\documentclass{iaus}
\usepackage{graphicx}

\title[S 275.~~BAL quasars] 
{Intermittent activity of radio sources. Accretion instabilities and
jet precession.}

\author[M. Kunert-Bajraszewska, A. Janiuk, A.Siemiginowska,
M.Gawro\'nski]   
{M. Kunert-Bajraszewska$^1$,
A. Janiuk$^2$, A.Siemiginowska$^3$ \and M.Gawro\'nski$^1$}

\affiliation{$^1$Toru\'n Centre for Astronomy, N. Copernicus University,
Gagarina 11, 87-100 Toru\'n, Poland\\ email: {\tt magda@astro.uni.torun.pl} \\[\affilskip]
$^2$Center for Theoretical Physics, PAN, Al.Lotnikow
32/46, 02-668 Warsaw, Poland\\[\affilskip]
$^3$Harvard Smithsonian Center for Astrophysics, 60 Garden St,
Cambridge, MA 02138}

\pubyear{2011}
\volume{275}  
\pagerange{119--126}
\jname{Jets at all Scales }
\editors{A.C. Editor, B.D. Editor \& C.E. Editor, eds.}
\begin{document}

\maketitle

\begin{abstract}
We consider the radiation pressure instability operating on short timescales
($10^{3} - 10^{6}$ years) in the accretion disk around a supermassive black hole
as the origin of the intermittent activity of radio sources.
We test whether this instability can be responsible for short ages
($<10^{4}$ years) of Compact Steep Spectrum sources 
measured by hot spots propagation 
velocities in VLBI observations and statistical overabundance
of Gigahertz Peaked Spectrum sources.The implied timescales are consistent
with the observed ages of the sources.
We aslo discuss possible implications of the intermittent activity on the 
complex morphology of radio sources, such as the quasar 1045+352, 
dominated by a knotty jet showing several bends. 
It is possible that we are whitnessing an ongoing 
jet precession in this source due to internal instabilities within the jet flow.

\end{abstract}

\firstsection
\section{Introduction}

The compact radio sources consist of two population of objects:   
the gigahertz-peaked spectrum (GPS) and compact steep spectrum (CSS) 
sources which are considered to be young and evolve into large radio objects, during their lifetimes 
(\cite[O'Dea,1998]{odea98}). However, it has already been pointed out by some authors
(\cite[Gugliucci et al.2005,
Ku\-nert-\-Baj\-ra\-szew\-ska et al.2006]{gug05,kun06})
that there exists a group of GPS/CSS sources that will never evolve to
become large scale objects, at least in a given cycle of activity if it is recurrent.
These sources can be named short-lived radio objects. 

We associated the existence of short-lived compact radio sources with the
intermittent activity of the central engine caused by a radiation pressure
instability within an accretion disk, which we briefly mention below.

\section{Discussion and Results}
According to the accretion disk instability model, for a given black 
hole mass, the larger the
mean accretion rate, the longer the duration of a cycle episode, both
in hot and cold states. In addition the viscosity parameter affects the
results and for smaller viscosity, the cycle duration is longer. This 
relation was calibrated by \cite{czerny} using a grid of models for  
various black hole masses and Eddington ratios.   
The outbursts are associated with the ejections of radio jets.  The
jets are then turned-off between the outbursts and each radio
structure will represent a new outburst. In case of an apparently
young, compact source we can suspect that in fact it is an old,  
reactivated object, in which the vast radio structures have already faded
away and are not visible.  
We have applied i.a. the accretion disk
instability model and jet precession model to explain the
potential reactivation of the
1045+352 core (\cite[Ku\-nert-Baj\-ra\-szew\-ska et al.2010]{kun10}). 

1045+352 is a CSS object and a HiBAL quasar at a medium redshift. Its
linear size ($\sim$4\,kpc) indcate it is a 
young object in the early phase of quasar evolution.
The radio morphology of 1045+352 is dominated by the
strong radio jet resolved into many sub-components and changing the  
orientation during propagation in the central regions of the host    
galaxy. As a consequence we observe at least three phases of jet activity
indicate different directions of the jet outflow: components $A_{2}$-$A_{3}$
as the oldest one, structure
$A_{1}$-B as the younger one, and the jet A as the current activity
direction (Fig.1).

\begin{figure}
\centering
\resizebox{10.5cm}{!}{\includegraphics{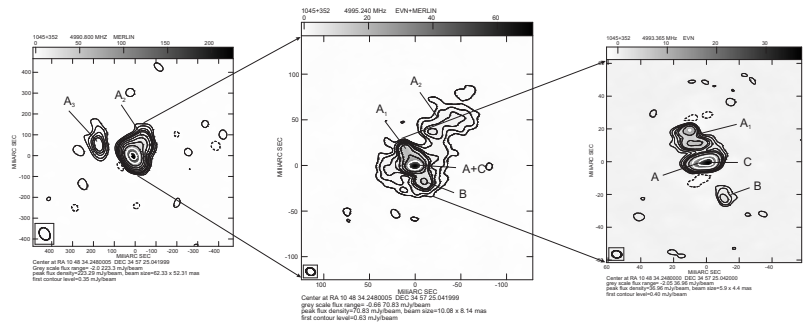}}
\caption[]{\footnotesize Radio images of 1045+352 at 5\,GHz made with (from left): MERLIN,
EVN+MERLIN, EVN. Contours increase by a
factor 2 and the first
contour level corresponds to ~3$\sigma$. Indications: C - radio core,
A-$A_{3}$ - radio jet, B - probable counter-jet.}
\end{figure}

The results of the applied accretion disk instability model (\cite[Janiuk et
al.2002, Czerny et al.2009]{jan02, czerny}) is not in
agreement with the 1045+352 structure and size:
(a) the calculated duration of the activity phase of the quasar is too short to
enable the source grow to the observed size, (b) it cannot reproduce
misalignment between the young and old radio structures. Such misalignment
could be the result of the changed direction of the jet axis between
the activity episodes - precession. In the case of 1045+352 we considered
the precession of the innermost accretion disk
due to internal instabilities within the accretion flow (\cite[Janiuk et
al.2008]{jan08}). This precession model reproduced well the observed complex
structure of 1045+352. However, as discussed by \cite{czerny}, the mechanism
of the accretion disk instability can be considered as the one that could
explain 
the apparent statistical excess of the compact radio sources with respect to the   
galaxies with extended radio structures.

\acknowledgements
\noindent
{\footnotesize This work was supported by the Polish Ministry of Science and   
Higher Education under grant N N203 303635.}

\end{document}